\documentclass[prl,aps, twocolumn]{revtex4}

\usepackage{graphicx, epsfig}
\usepackage{amssymb,amsmath}

        \def \o{\omega}
        
\def \e{\epsilon}       \def \th{\theta}
        
    \renewcommand{\o}{\omega}

   \def \S{\Sigma}    
\def \D{\Delta}

\def \h{\hbar}   

\def \del{\partial}    

\def\lbc{\left[}    \def\rbc{\right]}

\newcommand{\phm}{\phi_{\rm m}}
\newcommand{\phmt}{\phm(t)}

\newcommand{\dF}{\delta}
\newcommand{\dFf}{\delta_{\rm f}}
\newcommand{\dFi}{\delta_{\rm i}}
\newcommand{\dFc}{\delta_{\rm c}}

\begin{document}

\title{Sweeping a molecular Bose-Einstein condensate across a Feshbach resonance.}

\author{Masudul Haque}  \email{haque@phys.uu.nl}
\affiliation{Institute for Theoretical Physics, Utrecht
University, Leuvenlaan 4, 3584 CE Utrecht, The Netherlands}

\author{H. T. C. Stoof}  \email{stoof@phys.uu.nl}
\affiliation{Institute for Theoretical Physics, Utrecht
University, Leuvenlaan 4, 3584 CE Utrecht, The Netherlands}

\date{\today}


\begin{abstract}

We consider the dissociation of a molecular Bose-Einstein condensate
during a magnetic-field sweep through a Feshbach resonance that starts
on the molecular side of the resonance and ends on the atomic side. In
particular, we determine the energy distribution of the atoms produced
after the sweep. We find that the shape of the energy distribution
strongly depends on the rate of the magnetic-field sweep, in a manner
that is in good agreement with recent experiments.

\end{abstract}


\maketitle



{\it Introduction.} --- Feshbach resonances have established
themselves as one of the most important tools in the area of
ultracold atomic gases \cite{duine_thesis}. These resonances
provide the opportunity to make use of an applied magnetic field
to tune the interactions between the atoms, essentially by
changing the energy of a diatomic molecular state from a region
where the molecules are more stable to a region where the atoms
are more stable. For bosonic atoms, Feshbach resonances allow for
coherent oscillations between atomic and molecular Bose-Einstein
condensates \cite{Donley-etal_pulse_nature2002}. Moreover, in a
Fermi gas these Feshbach resonances recently have made a novel
BCS-BEC crossover phenomenon accessible to experiments
\cite{regal2004,MIT,Duke,Innsbruck,ENS}.

In this Letter, we study the response of a spatially uniform molecular
condensate to a magnetic-field sweep across a Feshbach resonance. The
essential physics of the nonequilibrium dissociation of the molecules
is the same for bosonic and fermionic atoms, except for the feedback
effects arising from the atomic statistics.  We show that under
appropriate conditions these feedback effects can be neglected, and in
that case our results are equally valid for fermionic and bosonic
atoms.

We focus in particular on two quantities. First, we develop a
method for determining the decay of the molecular condensate
during the sweep. Second, we investigate for different kinds of
sweeps the resulting energy spectrum of the dissociated atoms,
which turns out to be closely related to the time evolution of the
molecular condensate. Our approach to the time evolution is based
on the equation of motion for the molecular condensate wave
function $\phmt$, developed by Duine and Stoof
\cite{duine-stoof_J-OptB_2002,duine_thesis}. While no experiment
has yet probed the molecular dissociation as a function of time,
the decay of the molecular condensate density is in principle
experimentally accessible by a series of destructive measurements. 

Measurements of atomic energies after a dissociation ramp have been
performed recently by Mukaiyama \emph{et al}.\
\cite{ketterle_sweep-energy_nov03} and by D\"urr \emph{et al}.\
\cite{rempe_sweep-expt_may04}.  The ramps of Ref.\
\cite{ketterle_sweep-energy_nov03} are restricted to cases where all
the molecules dissociate during the sweep. The shape of the
dissociation spectrum for this type of sweep is well described by an
analytical expression, derived by Mukaiyama \emph{et al}.\
\cite{ketterle_sweep-energy_nov03} and also by G\"oral \emph{et al}.\
\cite{julienne_dec03}. For such a relatively slow type of ramp, our
calculated spectra agree with this analytical form as well. In Ref.\
\cite{rempe_sweep-expt_may04} also a different, more rapid, type of
sweep experiment is reported, in which most of the dissociation occurs
\emph{after} the sweep is over. As a result there is now a sharp peak
in the atomic energy distribution corresponding to half the molecular
energy at the end of the sweep. These spectra are not described by the
above mentioned analytical result. Our approach, however, allows us to
calculate also the spectra for such fast sweeps.

Both these sweep experiments involve bosonic atoms, in particular
$^{23}$Na and $^{87}$Rb, respectively. An additional motivation for
investigating sweeps in the molecule-to-atom direction comes from
considerations related to the cooling of fermionic atoms such as
$^6$Li and $^{40}$K. To avoid the
problems due to Fermi statistics, one proposal is to first cool
diatomic molecules to Bose-Einstein condensation, and then to slowly
sweep the system across the resonance to the atomic side. However, the
effect of the sweep rate on the resulting temperature of the atomic
gas is not well-understood at present.  It is, therefore, important to
study the dissociation process and the spectrum during a sweep across
a Feshbach resonance in some detail.

\emph{Generalized Gross-Pitaevskii equation.} --- A Feshbach
resonance is characterized by its magnetic field location $B_0$,
its width $\D{B}$, the background scattering length $a_{\rm bg}$,
and the magnetic-moment difference $\D\mu$ between two atoms and
the bare molecule associated with the Feshbach resonance.
Assuming that these quantities are for the atomic gas of interest
known from experiment, the equation of motion for the bare molecular
condensate wave function $\phm(t)$ becomes \cite{duine_thesis}
\begin{equation}  \label{eq_duine}
i\hbar\del_t \phm(t) ~=~ \lbc \dF(t) - i \eta \sqrt{i\hbar\del_t}
\rbc \phm(t)~,
\end{equation}
where $\dF(t)=\D\mu(B(t)-B_0)$ is the so-called detuning that is time
dependent during the sweep. The imaginary term in this equation of
motion describes the decay of molecules into pairs of atoms and is
determined in the Bose case by the quantity $\eta = m^{1/2}a_{\rm
bg}\D{B}\D\mu/\hbar = m^{3/2}g^2/2\pi\h^3$, where $g$ is the
atom-molecule coupling. When the atoms are fermionic, the right-hand
side is a factor of two smaller. Note that $\eta^2$ has the units of
energy and physically quantifies the width of the resonance. The
imaginary term arises from the molecular self energy
\begin{equation}     \label{eq_self-energy}
\hbar\S_{\rm m}(z) =
\begin{minipage}{3.5cm} \begin{center}
\epsfig{file=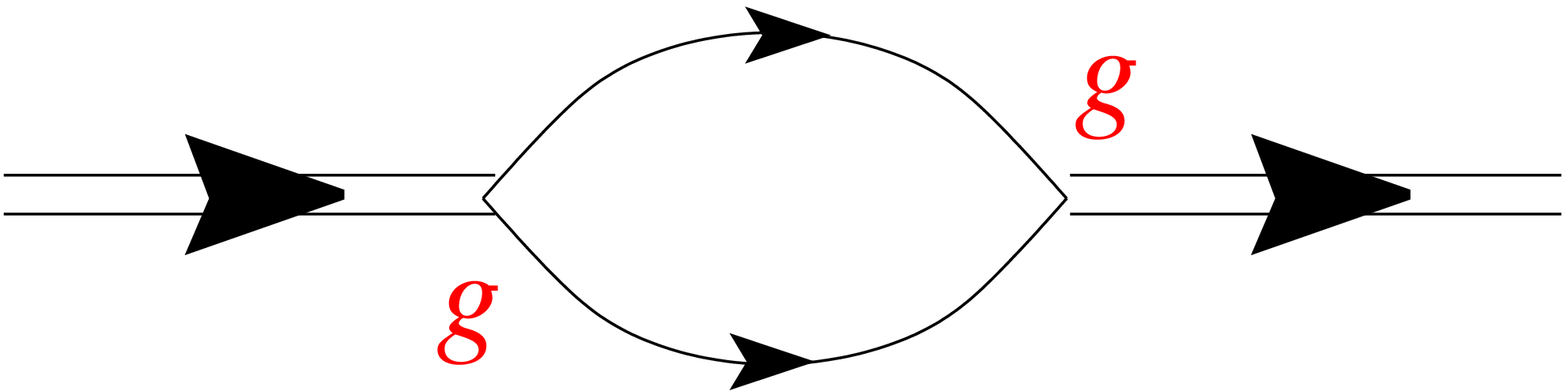,width=3cm}\end{center}\end{minipage}
= \eta \sqrt{-z}  \, ,
\end{equation}
where the single and double lines represent atomic and molecular
propagators, respectively. The square root of energy reflects the
three-dimensional final density of states expected from a Fermi's
Golden Rule calculation of the dissociation process. The
$\sqrt{i\hbar\del_t}$ form in Eq.~\eqref{eq_duine} arises from
evaluating the complex self energy in Eq.~\eqref{eq_self-energy}
from above on the real axis.

The above self energy leads physically to a dressing of the molecules
and as a result to a molecular density of states that for negative
detuning contains a delta function with weight $Z =
(1+\eta/2\sqrt{|\e_{\rm m}|})^{-1}$ at the dressed molecular binding
energy $\e_{\rm m}(\dF) = \dF-\eta^2/2 + \eta\sqrt{\eta^2/4-\dF}$ \cite{duine_thesis}.
Since the wave function $\phm$ appearing in Eq. \eqref{eq_duine}
represents the condensate wave function for bare molecules, the
dressed molecular density is a factor $Z^{-1}$ larger than the bare
molecular density and equal to $|\phm|^2/Z$.  
Moreover, the bare molecular density can now change in two
ways. Coherent changes involve a shift in the linear superposition
making up the dressed molecular wave function, i.e., a change of the
bare molecular fraction $Z$, while the dressed molecular density
remains constant \cite{duine_thesis, Donley-etal_pulse_nature2002}.
Incoherent changes are associated with the break-up of the dressed
molecule itself.

The generalized Gross-Pitaevskii equation in Eq.~\eqref{eq_duine}
involves some approximations. First, we are considering sufficiently
low molecular densities that the molecule-molecule interactions can be
neglected for most of the sweep duration, which is justified for
the experiments of interest to us. Second, we are not including the
effect of the thermal part of the molecular gas. This is justified for
sufficiently low temperatures. Most importantly, we are neglecting
feedback effects, i.e., the dynamics and statistics of the atoms that
are formed during and after the sweep. In particular, we do not allow
for their reassociation.  The neglect of atomic statistics, i.e., Bose
enhancement or Pauli blocking, is reasonable for monotonic sweeps,
because the dissociation then always occurs at different energies. In
that case we can estimate that the atomic statistics can be absolutely
neglected if the molecular density obeys $n_{\rm m} <
\hbar\dot{\dF}/\sqrt{2}\pi^2g^2$. Not surprisingly, this approximation
is better for faster ramps, and worst for a constant positive
detuning.

%

We now outline the method of solving Eq. \eqref{eq_duine}, given a
particular time-dependent detuning $\dF(t)$. We are especially
interested in magnetic-field sweeps across the resonance that take
the detuning $\dF$ from some negative value $\dFi$ at
$t=0$ to the final positive value $\dFf$ at time $t=T$.
Introducing an auxiliary constant detuning $\dFc$, we can write
\begin{equation}     \label{eq_dFc}
\lbc i \hbar\del_t + i\eta\ \sqrt{i\hbar\del_t} - \dFc \rbc
\phm(t) ~=~ \lbc \dF(t) - \dFc \rbc \phm(t)~.
\end{equation}
The solution of this differential equation is the sum of a
particular solution and the solution of the corresponding
homogeneous equation. We thus have
\begin{equation}  \label{eq_integral-eq}
\phm(t) = \psi(t) ~+~ 
\int_{-\infty}^t dt' \ G(t-t')\ \lbc \dF(t') - \dFc \rbc
\ \phm(t')   \,\, .
\end{equation}
Here $G(t)$ is the Green's function for the operator acting on
$\phmt$ in the left-hand side of Eq.~\eqref{eq_dFc}, i.e.,
\begin{equation}
G(t) ~=~ \int_{-\infty}^{\infty} \frac{d\o}{2\pi} \ \frac{
e^{-i{\o}t}}{\hbar\o^+ - \eta\sqrt{-\hbar\o^+}-\dFc}~,
\end{equation}
where $\hbar\o^+=\hbar\o+i0$. The real part of the pole in $G(\o)$
gives the molecular energy and the imaginary part gives the
dissociation rate, both at the detuning $\dFc$. Moreover, $\psi(t)$ is
the solution of the homogeneous equation of Eq.~(\ref{eq_dFc}). For
$\dFc<0$, the solution is an oscillatory function, $\psi(t)\propto
e^{-i{\o_{\rm c}}t/\hbar}$, with a frequency determined by the
molecular binding energy corresponding to the detuning $\dFc$, i.e.,
$\h\o_{\rm c} = \e_{\rm m}(\dFc)$.  In the following we use $\dFc=\dFi$,
because the integral in Eq. \eqref{eq_integral-eq} then has a lower
limit of $0$ instead of $-\infty$.

The retarded Green's function $G(t)$ is calculated by closing the
contour in the negative half-plane. The integrand has a branch cut
because of the square root in the energy denominator. For $\dFc<0$,
there is also a pole at $\e_{\rm m}$. Taking these into account we
obtain
\begin{multline*}
G(t) ~=~ \th(t)\, \lbc - \frac{i\eta}{\pi} \int_0^\infty d\o
\frac{\sqrt{\hbar\o}e^{-i{\o}t}}{(\hbar\o-\dFc)^2 + \eta^2\hbar\o}
\right.
\\
\left.
~-~  \frac{i}{\hbar} e^{-i\o_{\rm c}t/\hbar}\,
\frac{2\sqrt{-\hbar\o_{\rm c}}}{\eta + 2\sqrt{-\hbar\o_{\rm c}}}
\rbc.
\end{multline*}
Having calculated the Green's function $G(t)$ and the wave function
$\psi(t)$ corresponding to $\dFc$, we can numerically solve
Eq.~\eqref{eq_integral-eq} for $\phm(t)$. 

\begin{figure}
\resizebox{8.0cm}{!}{\includegraphics{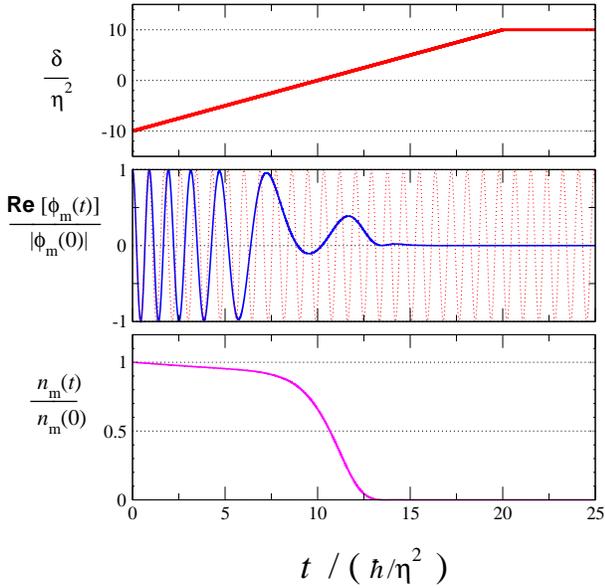}}
\caption{\label{fig_decay1}
  Decay of the molecular wave function
$\phm(t)/|\phm(0)|$ with time, when the detuning changes linearly with
time from $\dFi = -10\eta^2$ to $\dFf = +10\eta^2$ as shown in the top
panel. The center panel shows the real part of $\phm(t)$. The bottom
panel shows the decay of the molecular density, $n_{\rm m}(t)/n_{\rm
m}(0) = |\phm(t)|^2/|\phm(0)|^2$.}
\end{figure}

{\it Magnetic field sweeps.} --- In Fig.\ \ref{fig_decay1} we show the
decay of the bare molecular wave function $\phmt$ for a particular
ramp that is also shown in the top panel.  The center panel shows the
oscillatory decay of the real part of $\phmt$. The imaginary part
yields similar qualitative information and is not shown. The dotted
background is the nondecaying oscillatory $\phm(t)$ for a negative
detuning that is held fixed at $\dFi$. We see that $\phmt$ for the
ramp has a gradually decreasing oscillation frequency, corresponding
to the decreasing magnitude of the binding energy $\e_{\rm m}(\dF)$ as
$\dF$ approaches zero from the negative side.  The bottom panel shows
the decay of the bare density $n_{\rm m}(t) = |\phm(t)|^2$.  The decay
for $\dF(t)<0$ is mostly coherent, so that the dressed molecule
density remains almost fixed.  On the other hand, the decay at
positive detuning is incoherent, and the decay of the bare density
$n_{\rm m}(t)$ corresponds to the decay of dressed molecules.

The ramp in the experiment of Mukaiyama {\it et al}.\
\cite{ketterle_sweep-energy_nov03} involves an initial and final
detuning $\dF_{\rm i,f}$ of about $\mp 50\eta^2$, and ramp times of $T
\gtrsim 100 \h/\eta^2$. For the results shown in Fig.\
\ref{fig_decay1}, we used smaller values for both, so as to have fewer
oscillations to visualize.
\vspace{0.3cm}

We now outline how the incoherent dissociation spectrum is obtained,
once $\phmt$ has been calculated. We calculate the time derivative of
the molecular density from the equation of motion in
Eq.~\eqref{eq_duine}. Not surprisingly, only positive frequencies
contribute, because the molecules are stable for negative
energies. We obtain
%
%
%
\begin{equation}  
\left.
\int_{-\infty}^{\infty} dt\ \dot{n}_{\rm m} \right|_{\rm incoh}
~=~ -\frac{\eta}{\pi \h}
\int_0^\infty d\o \sqrt{\hbar\o} \left|
 \phm(\o)\right|^2   \,\, ,
\end{equation}
where $\phm(\o)$ is the Fourier transform of $\phm(t)$.  The integrand
in the right-hand side is interpreted as the molecular dissociation
spectrum $f_{\rm mol}(\e) = \eta \sqrt{\e}
\left|\phm(\e)\right|^2/\pi\h$.
Since each atom has half the energy of the dissociating molecule,
$f_{\rm mol}(\e)$ corresponds to the atomic spectrum at $\e/2$, i.e.,
$f_{\rm at}(\e) = 4 f_{\rm mol}(2\e)$. The additional factor of 2 is
because each molecule gives rise to two atoms.

Before presenting results for the spectrum, we briefly discuss the
analytical result mentioned previously \cite{julienne_dec03,
ketterle_sweep-energy_nov03}. This result can be derived by assuming
that for a particular detuning, each decaying molecule has a sharply
defined energy. This means that the width of the molecular spectral
function is neglected. Moreover, it is implicitly assumed that all the
molecules are dissociated before the sweep is over.
If we denote by $n_{\rm m}(\e)$ the density of molecules at energy $\e$
and use $dt =
d\e/\dot{\e} \simeq d\e/\dot{\dF}$, we have $n_{\rm m}(\e+d\e) -
n_{\rm m}(\e) = -(2\eta\sqrt{\e}/\hbar\dot{\dF}){n_{\rm m}(\e)}d\e$.  
Solving for $n_{\rm m}(\e)$, the molecular dissociation spectrum
follows then from $f_{\rm mol} = dn_{\rm m}(\e)/d\e$. The atomic
energy spectrum $f_{\rm at}$ can be deduced from $f_{\rm mol}$ as
before with the result
\begin{equation}   \label{eq_JK}
f_{\rm at}(\e) ~=~ n_{\rm m}(0) \frac{8\eta\sqrt{2\e}}{\hbar\dot{\dF}} \,
\exp\lbc -\, \frac{4\eta\,(2\e)^{3/2}}{3\hbar\dot{\dF}} \rbc~.
\end{equation}

\begin{figure}
\resizebox{8.0cm}{!}{\includegraphics{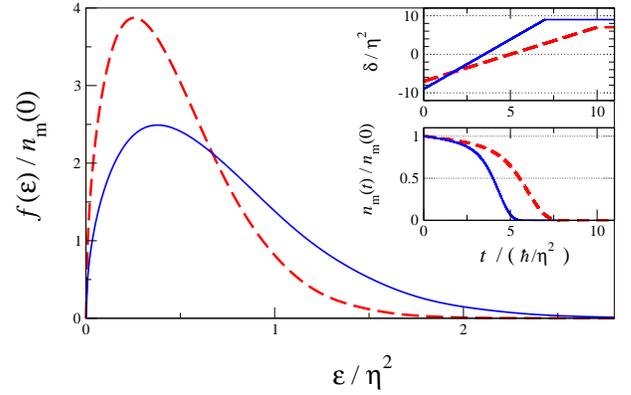}}
\caption{   \label{fig_spectra_JK-type}
  Spectra for cases where the analytic derivation is
valid, because the molecules decay almost completely before the sweep
is over.  Insets show the detuning ramps with different ramp rates,
and the bare molecular condensate densities. }
\end{figure}

We stress that this expression is not a Landau-Zener or adiabatic
result. Its derivation does not require the sweep to be slow, although
it is required that the sweep continues until all the molecules have
dissociated. Moreover, the derivation relies on the approximation $\e
\simeq \dF$, which is not accurate at small energies $\e \ll \eta^2$,
for which we have that $\e \simeq \dF^2/\eta^2$.  In
Fig.~\ref{fig_spectra_JK-type} we show calculated spectra for two
cases where the above mentioned conditions holds.  The spectra indeed
have the exponential form of Eq.~\eqref{eq_JK}, however they are in
general normalized according to the initial dressed molecular density
$n_{\rm m}(0)/Z_{\rm i}$ instead of the initial bare molecular density
$n_{\rm m}(0)$.  The ramp represented by the solid curves has a higher
ramp rate and hence the resulting spectrum is tilted to higher
energies. Note that experimentally, only the total atomic energy has
been measured as a function of ramp speeds
\cite{ketterle_sweep-energy_nov03, rempe_sweep-expt_may04}. These
measured total energies are fit well by the total energies calculated
analytically as $\int{d\e}\ {\e}\ f_{\rm at}(\e)$. Since our spectra
match the analytic result for the experimentally relevant case of
$Z_{\rm i} \simeq 1$, agreement with the measured total energies
follows, and is not shown here.

\vspace{0.3cm}

\begin{figure}
\resizebox{8.0cm}{!}{\includegraphics{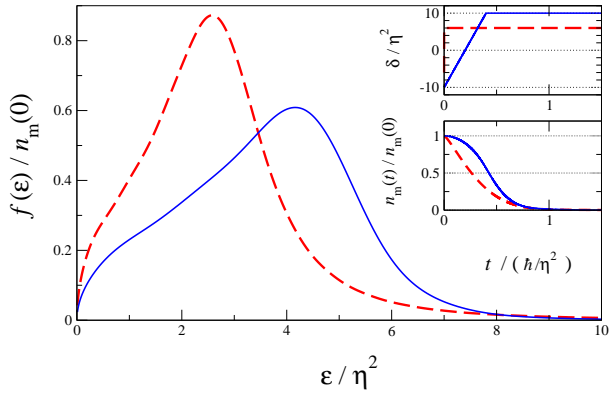}}
\caption{   \label{fig_spectra_non-JK-type}
  Spectra for cases where the analytic form is not
valid. Dashed curves: Infinitely fast ramp at which the detuning
switched from $\dFi = -6\eta^2$ to $\dFf = +6\eta^2$ at $t=0$.
Full curves: Fast ramp that is over before all the molecules can
decay. }
\end{figure}

In Fig.~\ref{fig_spectra_non-JK-type} we show spectra calculated for
two cases where Eq.~\eqref{eq_JK} does not hold because the sweep is
over before all the molecules have decayed. This type of sweep has
been employed in a recent experiment
\cite{rempe_sweep-expt_may04}. The density images in that experiment
indicate that the spectrum is peaked at the energy corresponding to
the final detuning. This feature is clear in
Fig.~\ref{fig_spectra_non-JK-type}. The dashed curves represent an
infinitely fast ramp. The spectrum shows a peak around $\dF_{\rm
f}/2$, but is spread over a relatively large range of energies.  The
spread demonstrates that our method takes into account the
distribution of molecular energies, i.e., the width of the molecular
density of states at a fixed positive detuning. The solid curves in
Fig.~\ref{fig_spectra_non-JK-type} represents a ramp that is over
before all the molecules have decayed. A significant fraction decays
while the detuning is constant at its final value. The atomic spectrum
here also shows a peak at an energy around half the final detuning.
\vspace{0.3cm}

\begin{figure}
\resizebox{8.0cm}{!}{\includegraphics{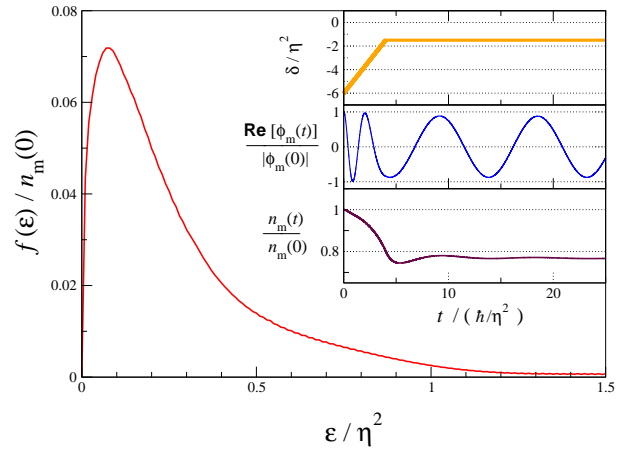}}
\caption{   \label{fig_neg-to-neg}
  A sweep that starts and ends at negative detuning. In
this case, the final density is expected to be $(Z_{\rm f}/Z_{\rm
i})n_{\rm m}(0) \simeq 0.7775 n_{\rm m}(0)$.  The actual final
density, $n_{\rm f} \simeq 0.735 n_{\rm m}(0)$, is smaller, in
agreement with the nonzero atomic spectrum.  }
\end{figure}

In order to demonstrate mainly coherent loss of bare molecules, we
display in Fig.~\ref{fig_neg-to-neg} a sweep that ends at negative
detuning, i.e., before reaching the resonance.  The density of bare
molecules decreases, but most of this decrease is due to the coherent
change of the dressed molecular wave function.  The final density
$n_{\rm f}$ is slightly smaller than $(Z_{\rm f}/Z_{\rm i})n_{\rm
m}(0)$, where $Z_{\rm i}$ and $Z_{\rm i}$ are the weight of the delta
function in the molecular density of states at the initial and the
final detuning, respectively.  The difference between $n_{\rm f}$ and
$(Z_{\rm f}/Z_{\rm i})n_{\rm m}(0)$ appears as the total weight of the
incoherent spectrum.  Note also that the density undergoes some
coherent oscillations before settling at its final value.  These are
internal Josephson oscillations of the dressed molecular condensate.

\vspace{0.3cm}

{\it Conclusions.} --- We have shown how to calculate the time
evolution of the molecular condensate wave function $\phmt$ during
sweeps across a Feshbach resonance.  We have also shown how the time
dependence of this wave function leads to the dissociation spectrum.
Our calculation for the atomic energy spectrum is more general than
the result obtained by Mukaiyama \emph{et al}.\
\cite{ketterle_sweep-energy_nov03} and by G\"oral \emph{et al}.\
\cite{julienne_dec03}. The theory presented here is applicable for
ramps of essentially arbitrary shape and initial conditions. This has
been demonstrated with the fast ramps, in
Fig.~\ref{fig_spectra_non-JK-type}. Moreover, we have obtained the, in
first instance surprising, result that a nonzero dissociation spectrum
can also be obtained by a changing magnetic field that always remains
below the Feshbach resonance.

Finally, we point out two important issues that remain unsolved.
First is the problem of incorporating the effect of the atomic
statistics. In principle, this can be achieved by coupling the
generalized Gross-Pitaevskii equation for the molecular condensate
to a quantum Boltzmann equation for the atoms. It remains to be
seen, however, if this leads to a feasible approach. Second, our
calculation is based on the approximation for the molecular self
energy given in Eq.~\eqref{eq_self-energy}. For broad Feshbach
resonances, such as the one in $^6$Li that is widely used at
present for the study of the BCS-BEC crossover
\cite{MIT,Duke,Innsbruck,ENS}, the molecular self energy can no
longer be expressed in terms of the $\eta$ parameter alone, and
depends also on the background scattering length $a_{\rm bg}$
\cite{duine_thesis}. Incorporating this improvement is also left
for future work.

We thank Randy Hulet for helpful comments.  This work is supported by
the Stichting voor Fundamenteel Onderzoek der Materie (FOM) and the
Nederlandse Organisatie voor Wetenschaplijk Onderzoek (NWO).

\end{document}